\documentstyle[epsf,12pt]{article}
\begin{document}
\begin{center}
\Large{\bf A Heavy Ion Fireball freeze-out Dipion Cocktail for Au-Au Collisions at $\sqrt{s_{NN}}$=200 GeV $p_t$ dependence(Part 2).}\\
\large{R.S. Longacre$^a$\\
$^a$Brookhaven National Laboratory, Upton, NY 11973, USA}
\end{center}
 
\begin{abstract}
In this paper we use methods developed in Part 1 of The Dipion Cocktail, to fit
the $p_t$ dependence of dipions for mid-central Au-Au collisions at 
$\sqrt{s_{NN}}$=200 GeV. For the minijet fragmentation part we use PYTHIA
fragmentation as described in Part 1. For the thermal resonance production we
use an exponential growth behavior. The interference between the direct 
production of dipion pairs from non-resonance minijet fragmentation and 
re-scattering through resonance states gives a measure of the size of the 
re-scattering region. This size is contain in the $\alpha$ parameter of Part1.
We assumed a relationship between the $\alpha$ parameter and the mass shift of 
the $\rho$ resonance. The data used for the fits comes from the RHIC collider
as measured in the STAR experiment.
\end{abstract}
 
\section{Introduction} 

The ultra-relativistic heavy ion collision starts out as a state of high 
density nuclear matter called the Quark Gluon Plasma(QGP) and expands rapidly 
to freeze-out. During the freeze-out phase quarks and gluons form a system of 
strongly interacting hadrons. These hadrons continue to expand in a thermal 
manner until no further scattering is possible because the system becomes to 
dilute. However this transition from quarks and gluons(partons) into hadrons is
not a smooth affair. The expansion is very rapid and some faster or hard 
scattered partons fragment directly into hadron through a minijet\cite{Trainor}
process. Thus we have thermal and minijet hadrons present in the last 
scattering of the hadrons. The Dipion Cocktail Part 1 considered this mixture 
of sources and applied it to the dipion mass spectrum of the heavy ion fireball
formed in Au-Au collisions at $\sqrt{s_{NN}}$=200. Part 1 showed that both 
thermal or soft production of hadrons and the minijet fragmented hadrons can
be described through a set of unified formal equations. Part 2(this paper) 
applies this formalism to the $p_t$ dependence of dipions for Au-Au collisions 
at $\sqrt{s_{NN}} = $ 200 GeV and 40\% to 80\% centrality.

The paper is organized in the following manner:

Sec. 1 Introduction. Sec. 2 Review of the two component model which we use to 
fit the dipion data within a set of $p_t$ ranges. Sec. 3 Discussion of the 
relationship between the $\alpha$ parameter and the mass and widths of 
resonances. Sec. 4 we present a fit to 19 $p_t$ ranges for Au-Au collisions at 
$\sqrt{s_{NN}} = $ 200 GeV and 40\% to 80\% centrality. Sec. 5 Summary and 
Discussion.

\section{Two component model with Breit-Wigner parameters}

In this section we will alter equation 6 of Part 1 so it can use Breit-Wigner 
parameters (mass, width) instead of phase shifts. We will also need to
modify the re-scattering part of the equation in order to have the correct
threshold behavior we have introduced in Part 1 for the  minijet partial waves.
The phase shift can be written for the $\ell^{th}$ wave as
\begin{equation}
cot\delta_\ell = \frac{(M_{\ell}^2 - M_{\pi\pi}^2)}{M_{\ell}\Gamma_{\ell}},
\end{equation}
where $M_{\ell}$ is the mass of the resonance in the $\ell$wave and 
$\Gamma_{\ell}$ is its total width.
\begin{equation}
\Gamma_{\ell} = \Gamma_{0\ell} {\frac{qB_{\ell}(q/q_s)}{M_{\pi \pi}}\over{\frac{q_{\ell}B_{\ell}(q_{\ell}/q_s)}{M_{\ell}}}}
\end{equation}
with $\Gamma_{0\ell}$ the total width at resonance, $B_{\ell}$ is the
Blatt-Weisskopf-barrier factor\cite{Hippel} for the $\ell$ of the resonance, 
$q$ is the $\pi\pi$ center mass momentum, $q_{\ell}$ is $q$ at resonance, 
$M_{\ell}$ is the mass of the resonance, and $q_s$ is center mass momentum 
related to the size(1.0 fm is used $q_s$ = .200 GeV/c).

Using equation 1 we rewrite equation 6 of Part 1 as
\begin{equation}
|T_{\ell}|^2  = |D_{\ell}|^2 \frac{sin^2\delta_{\ell}}{PS_{\ell}} +
\frac{|A_{\ell}|^2sin^2\delta_{\ell}}{PS_{\ell}} \left| \alpha + PS_{\ell} cot\delta_\ell \right|^2
\end{equation}

The $D_{\ell}$ is the thermal production term and is constant except for the 
Boltzmann weight(see equation 13 in Part 1). The expected threshold behavior 
$q^{2\ell+1}$ comes from the $sin\delta_{\ell}$ term. Since there is  
$sin^2\delta_{\ell}$ one of the $q^{2\ell+1}$ is killed off by dividing by 
$PS_{\ell}$. In Figure 6 of Part 1 we see we have put into our minijet 
$A_{\ell}$ the correct threshold $q^{2\ell+1}$ so we need to kill off the 
$q^{2\ell+1}$ of the other $sin\delta_{\ell}$ term. Therefore the above 
equation for our minijet $A_{\ell}$ we will use
\begin{equation}
|T_{\ell}|^2  = |D_{\ell}|^2 \frac{sin^2\delta_{\ell}}{PS_{\ell}} +
\frac{|A_{\ell}|^2sin^2\delta_{\ell}}{PS_{\ell}^2} \left| \alpha + PS_{\ell} cot\delta_\ell \right|^2
\end{equation}

Rewriting equation 6 of Part 1 for each partial wave with Breit-Wigner 
parameters the first term becomes

\begin{equation}
|T_{\ell}|_1^2  = |D_{\ell}|^2 \frac{M_{\pi\pi}^2}{\sqrt{M_{\pi\pi}^2 + p^2_t}} exp \frac{-\sqrt{M_{\pi\pi}^2 + p^2_t}}{T} \frac{M_{\ell}\Gamma_{\ell}}{(M_{\ell}^2 -M_{\pi\pi}^2)^2 + M_{\ell}^2\Gamma_{\ell}^2},
\end{equation}
while the second term
\begin{equation}
|T_{\ell}|_2^2  = |A_{\ell}|^2 \frac{M_{\ell}^2\Gamma_{\ell}^2}{(M_{\ell}^2 -M_{\pi\pi}^2)^2 + M_{\ell}^2\Gamma_{\ell}^2}\left|\alpha + \frac{2qB_{\ell}(\frac{q}{q_s})(M_{\ell}^2 -M_{\pi\pi}^2)}{M_{\pi\pi}M_{\ell}\Gamma_{\ell}}\right|^2 \left(\frac{M_{\pi\pi}^2}{4q^2B_{\ell}^2(\frac{q}{q_s})}\right).
\end{equation}

\begin{equation}
|T|^2 = \sum_{\ell} |T_{\ell}|^2
\end{equation}
where
\begin{equation}
|T_{\ell}|^2 = |T_{\ell}|_1^2 + |T_{\ell}|_2^2
\end{equation}
and $|A_0|^2 = S(M_{\pi^+\pi^-})$,$|A_1|^2 = P(M_{\pi^+\pi^-})$,$|A_2|^2 = D(M_{\pi^+\pi^-})$, and $|A_3|^2 = F(M_{\pi^+\pi^-})$. S, P, D and F comes from
subsection 5.2 of Part 1.

\section{A Relationship between the $\alpha$ parameter and the mass and widths of resonances.}

Equation 6 of Part 1 has an important factor the coefficient $\alpha$.
This coefficient is related to the real part of the $\pi \pi$ re-scattering
loop and is given by equation 9. When the pions re-scatter or interact at
a close distance or a point the real part of the loop $\alpha$ has its maximum 
value of $\alpha_0$. While if the pions re-scatter or interact at a distance 
determined by the diffractive limit the value of $\alpha$ is zero. The $\alpha$
which is the real part of the re-scattering factor has a simple form given by
\begin{equation}
\alpha = (1.0 - \frac{r^2}{r_0^2}) \alpha_0
\end{equation}
where $r$ is the radius of re-scattering in fm's and $r_0$ is 1.0
fm or the limiting range of the strong interaction ranging to  $r$ = 0.0 for 
point like interactions.

When $\pi \pi$ pairs interact at the diffractive limit their phase shift should
be the same as the phase shift of the vacuum. The same statement is true for
$\pi \pi$ interacting at a point since asymptotic freedom demands that the 
strong interaction should have no effect. However values of $\alpha$ in between
zero and $\alpha_0$ represent a confined volume where strongly interacting 
gluons, quarks and virtual mesons may influence the phase shift of the 
$\pi \pi$ system. Phase shifts under a Breit-Wigner parameters assumption
depend on the mass and width of the resonance parameter. In the next
section we use fits to data to determine the relationship between Breit-Wigner 
parameters and the value of $\alpha$. 

\section{STAR data dipion $p_t$ range (0.2 GeV/c $<$ $p_t$ $<$ 4.0 GeV/c)}

We have fitted 19 dipion $p_t$ ranges(see Table I) using equation 7 above for 
the STAR Au-Au collisions at $\sqrt{s_{NN}} = $ 200 GeV 40\% to 80\% 
centrality data. We included minijets up to $\ell$ = 3 and resonances $\sigma$ 
$\ell$ = 0, $\rho(770)$ $\ell$ = 1, and $f_2(1270)$ $\ell$ = 2. Using the 
arguments of Sec. 3 of Part 1, we added the $f_0$ as a direct thermal term 
($|T_0|_1^2$) and only the $\sigma$ interfered with $\ell$ = 0 minijet 
background. Two other thermal terms are present in the cocktail, the $k^0_s$ 
and the $\omega_0$. All the thermal terms have an exponential behavior with
dipion $p_t$. The spectrum of the minijet partial waves is obtained from
PYTHIA\cite{pythia}(see Sec. 5.2 of Part 1). We let the data determine which
minijet partial wave to add. We find only Swave minijet background is important
until $p_t$ equal to 1.1 GeV/c. Above 1.5 GeV/c all four minijet partial waves
are used up to Fwave. It should be noted Dwave and Fwave are small effects. 
We used PDG\cite{PDG} for the $f_2(1270)$ mass = 1.275 GeV and width = 
.185 GeV. The $f_0$ was fitted obtaining mass = 0.9727 $\pm$ .0039 GeV and
width = 0.04512 $\pm$ 0.01128 GeV. The $\sigma$ mass and width used was fixed
because it was ill determined. The mass used was mass = 1.011 GeV and width =
1.015 GeV. The $\rho$ mass and width is explained below.

Finally the threshold effective mass region .280 GeV to .430 GeV is dominated 
by the Swave and receives contributions from minijet fragmentation, $\pi \pi$ 
Swave phase shift, $\eta$ decay, HBT adding to the like sign $\pi \pi$ 
distribution that has been subtracted away from the unlike sign $\pi \pi$ and 
the coulomb correction between the charged pions. The minijet fragmentation is 
the least known of the effects since we relied on PYTHIA, however there are 
large uncertainty in all the other effects. So for these fits we let the 
minijet fragmentation be free to fit the data and let the Breit-Wigner 
parameters for the $\sigma$ determine the Swave phase shifts plus leaving 
out all other effects. 

For the $\alpha$ parameter in $p_t$ bins up to 1.1 GeV/c the minijet Swave 
interference is the determining factor. Above 1.1 GeV/c the Pwave interference
becomes most important. The values of $\alpha$ which gives a reasonable fit are
shown in Table II.

We have determined that the $\sigma$ pole or Breit-Wigner parameters is so far
away from the real axis thus it is too short lived to be influenced by hadronic
interactions. The $\rho$ phase shift being of a life time comparable to 
hadronic interaction taking place becomes most sensitive. We have found as a
function of $\alpha$ the best of $\rho$ width is always 0.147 GeV with an error
of $\pm$ .007 GeV. The mass however decreases as $\alpha$ grows, reaching a 
minimum of 0.738 GeV at an $\alpha$ of 0.907. This is a mass shift of 37 MeV. 
An $\alpha$ of 0.504 is the smallest $\alpha$ we find in our fits. A mass of 
0.775 GeV is the best fit when the value of $\alpha$ is at 0.504. Using 
equation 9 in Table II we determine the radius of $\pi \pi$ re-scattering for 
each $p_t$ bin. The value of $\alpha_0$ used in Table II is equal to 2.0 as 
determined in Appendix B of Part 1. Table II shows an interesting density 
effect around dipion $p_t$ of 0.6 to 1.0 GeV/c. If one consider that $p_t$ 
maybe related to fireball size through the idea of hubble flow, then pions with
a $p_t$ of around 0.4 GeV/c maybe coming from a less dense region in the
central part of the fireball. This could be a density wave effect. 

\clearpage

\bf Table I. \rm The $p_t$ ranges bins we fit.

\begin{center}
\begin{tabular}{|r|r|r|}\hline
\multicolumn{3}{|c|}{Table I}\\ \hline
$p_t$ bin number & lower edge(GeV/c) & upper edge(GeV/c) \\ \hline
1 & 0.2 & 0.4 \\ \hline
2 & 0.4 & 0.6 \\ \hline
3 & 0.6 & 0.8 \\ \hline
4 & 0.8 & 1.0 \\ \hline
5 & 1.0 & 1.2 \\ \hline
6 & 1.2 & 1.4 \\ \hline
7 & 1.4 & 1.6 \\ \hline
8 & 1.6 & 1.8 \\ \hline
9 & 1.8 & 2.0 \\ \hline
10 & 2.0 & 2.2 \\ \hline
11 & 2.2 & 2.4 \\ \hline
12 & 2.4 & 2.6 \\ \hline
13 & 2.6 & 2.8 \\ \hline
14 & 2.8 & 3.0 \\ \hline
15 & 3.0 & 3.2 \\ \hline
16 & 3.2 & 3.4 \\ \hline
17 & 3.4 & 3.6 \\ \hline
18 & 3.6 & 3.8 \\ \hline
19 & 3.8 & 4.0 \\ \hline
\end{tabular}
\end{center}
 
\clearpage

\bf Table II. \rm The $\alpha$ value, $\rho$ mass value and radius in each $p_t$ range.

\begin{center}
\begin{tabular}{|r|r|r|r|}\hline
\multicolumn{4}{|c|}{Table II}\\ \hline
$p_t$(GeV/c) & $\alpha$ & $\rho$ mass(GeV) & radius(f)\\ \hline
0.3 & 0.907 $\pm$ .028  & 0.738 $\pm$ .004 & .739 $\pm$ .010 \\ \hline
0.5 & 0.806 $\pm$ .025  & 0.748 $\pm$ .003 & .773 $\pm$ .008 \\ \hline
0.7 & 0.706 $\pm$ .022  & 0.755 $\pm$ .002 & .804 $\pm$ .007 \\ \hline
0.9 & 0.706 $\pm$ .022  & 0.755 $\pm$ .002 & .804 $\pm$ .007 \\ \hline
1.1 & 0.806 $\pm$ .025  & 0.748 $\pm$ .003 & .773 $\pm$ .008 \\ \hline
1.3 & 0.806 $\pm$ .025  & 0.748 $\pm$ .003 & .773 $\pm$ .008 \\ \hline
1.5 & 0.806 $\pm$ .025  & 0.748 $\pm$ .003 & .773 $\pm$ .008 \\ \hline
1.7 & 0.806 $\pm$ .025  & 0.748 $\pm$ .003 & .773 $\pm$ .008 \\ \hline
1.9 & 0.806 $\pm$ .025  & 0.748 $\pm$ .003 & .773 $\pm$ .008 \\ \hline
2.1 & 0.605 $\pm$ .019  & 0.759 $\pm$ .002 & .835 $\pm$ .006 \\ \hline
2.3 & 0.504 $\pm$ .016  & 0.775 $\pm$ .001 & .865 $\pm$ .004 \\ \hline
2.5 & 0.504 $\pm$ .016  & 0.775 $\pm$ .001 & .865 $\pm$ .004 \\ \hline
2.7 & 0.504 $\pm$ .016  & 0.775 $\pm$ .001 & .865 $\pm$ .004 \\ \hline
2.9 & 0.504 $\pm$ .016  & 0.775 $\pm$ .001 & .865 $\pm$ .004 \\ \hline
3.1 & 0.504 $\pm$ .016  & 0.775 $\pm$ .001 & .865 $\pm$ .004 \\ \hline
3.3 & 0.504 $\pm$ .016  & 0.775 $\pm$ .001 & .865 $\pm$ .004 \\ \hline
3.5 & 0.504 $\pm$ .016  & 0.775 $\pm$ .001 & .865 $\pm$ .004 \\ \hline
3.7 & 0.504 $\pm$ .016  & 0.775 $\pm$ .001 & .865 $\pm$ .004 \\ \hline
3.9 & 0.504 $\pm$ .016  & 0.775 $\pm$ .001 & .865 $\pm$ .004 \\ \hline
\end{tabular}
\end{center}

The 19 dipion spectrum for the $p_t$ bins are shown in Figure 1 through Figure 
19. 

\begin{figure}
\begin{center}
\mbox{
   \epsfysize 6.8in
   \epsfbox{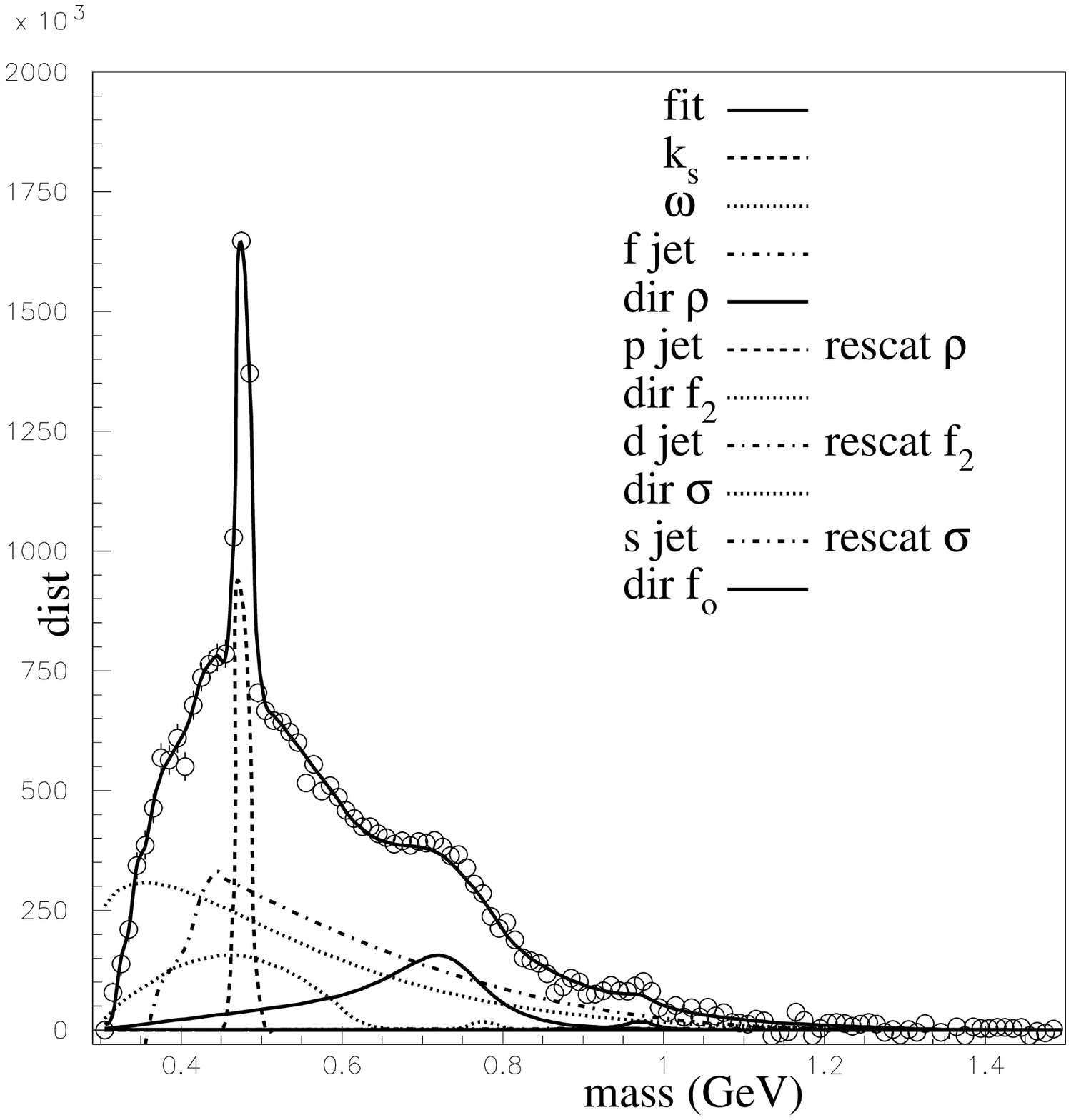}}
\end{center}
\vspace{2pt}
\caption{Fit to STAR dipion effective mass distribution (0.2 GeV/c $<$ $p_t$ 
$<$ 0.4 GeV/c) for Au-Au collisions at $\sqrt{s_{NN}} = $ 200 GeV 40\% to 80\% 
centrality using equation 7. See text for complete information.}
\label{fig1}
\end{figure}

\begin{figure}
\begin{center}
\mbox{
   \epsfysize 6.8in
   \epsfbox{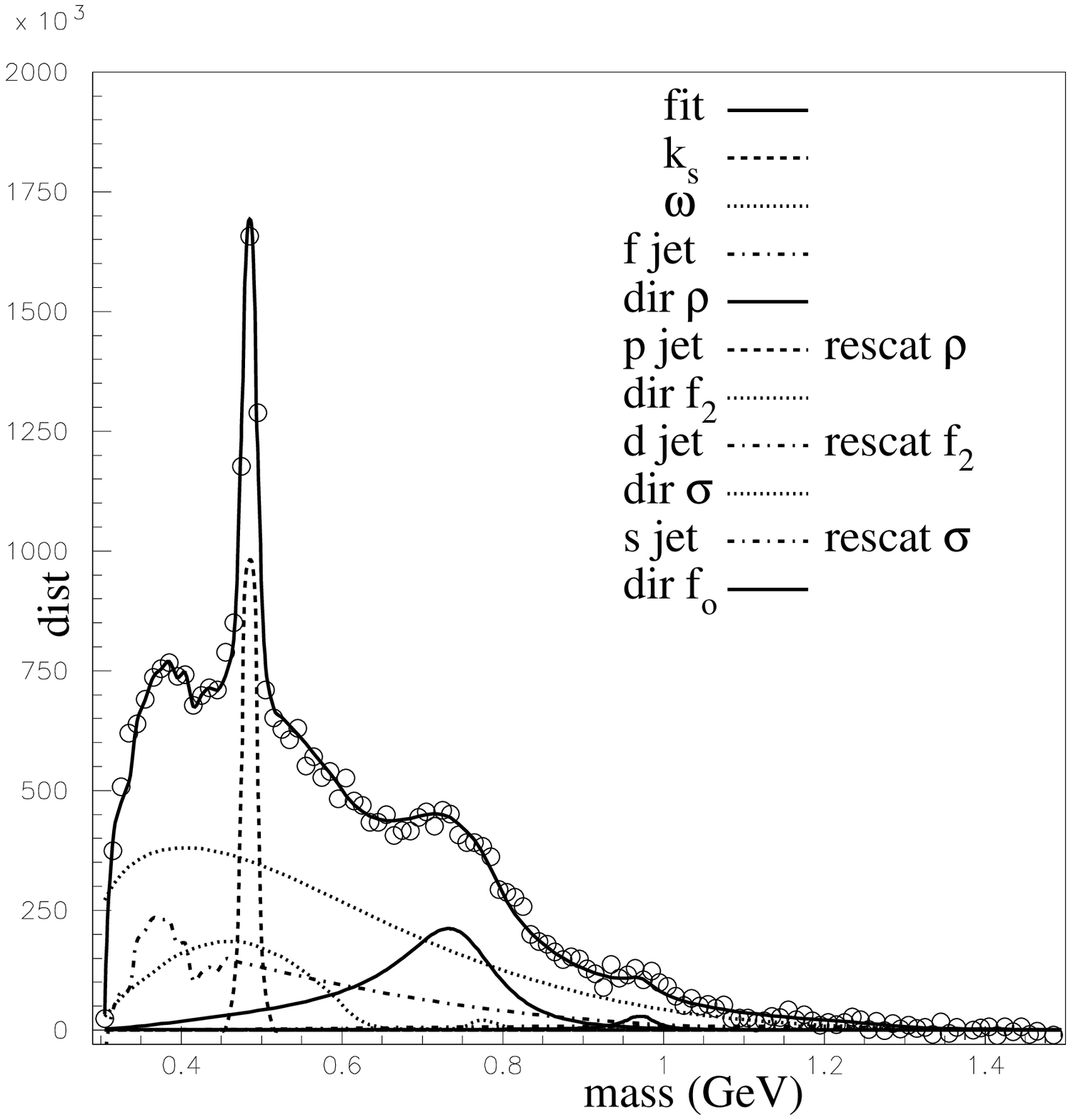}}
\end{center}
\vspace{2pt}
\caption{Fit to STAR dipion effective mass distribution (0.4 GeV/c $<$ $p_t$ 
$<$ 0.6 GeV/c) for Au-Au collisions at $\sqrt{s_{NN}} = $ 200 GeV 40\% to 80\% 
centrality using equation 7. See text for complete information.}
\label{fig2}
\end{figure}

\begin{figure}
\begin{center}
\mbox{
   \epsfysize 6.8in
   \epsfbox{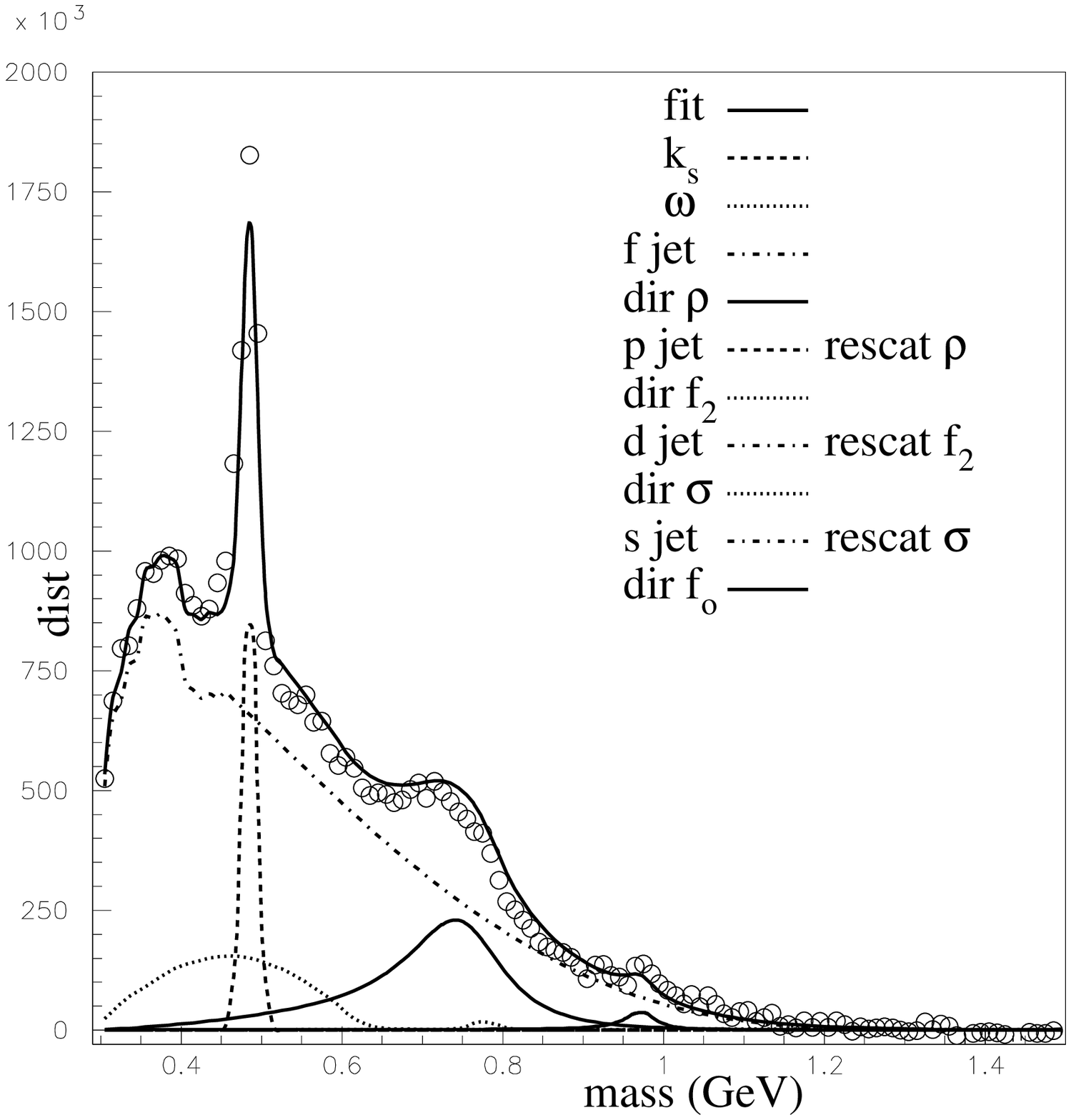}}
\end{center}
\vspace{2pt}
\caption{Fit to STAR dipion effective mass distribution (0.6 GeV/c $<$ $p_t$ 
$<$ 0.8 GeV/c) for Au-Au collisions at $\sqrt{s_{NN}} = $ 200 GeV 40\% to 80\% 
centrality using equation 7. See text for complete information.}
\label{fig3}
\end{figure}

\begin{figure}
\begin{center}
\mbox{
   \epsfysize 6.8in
   \epsfbox{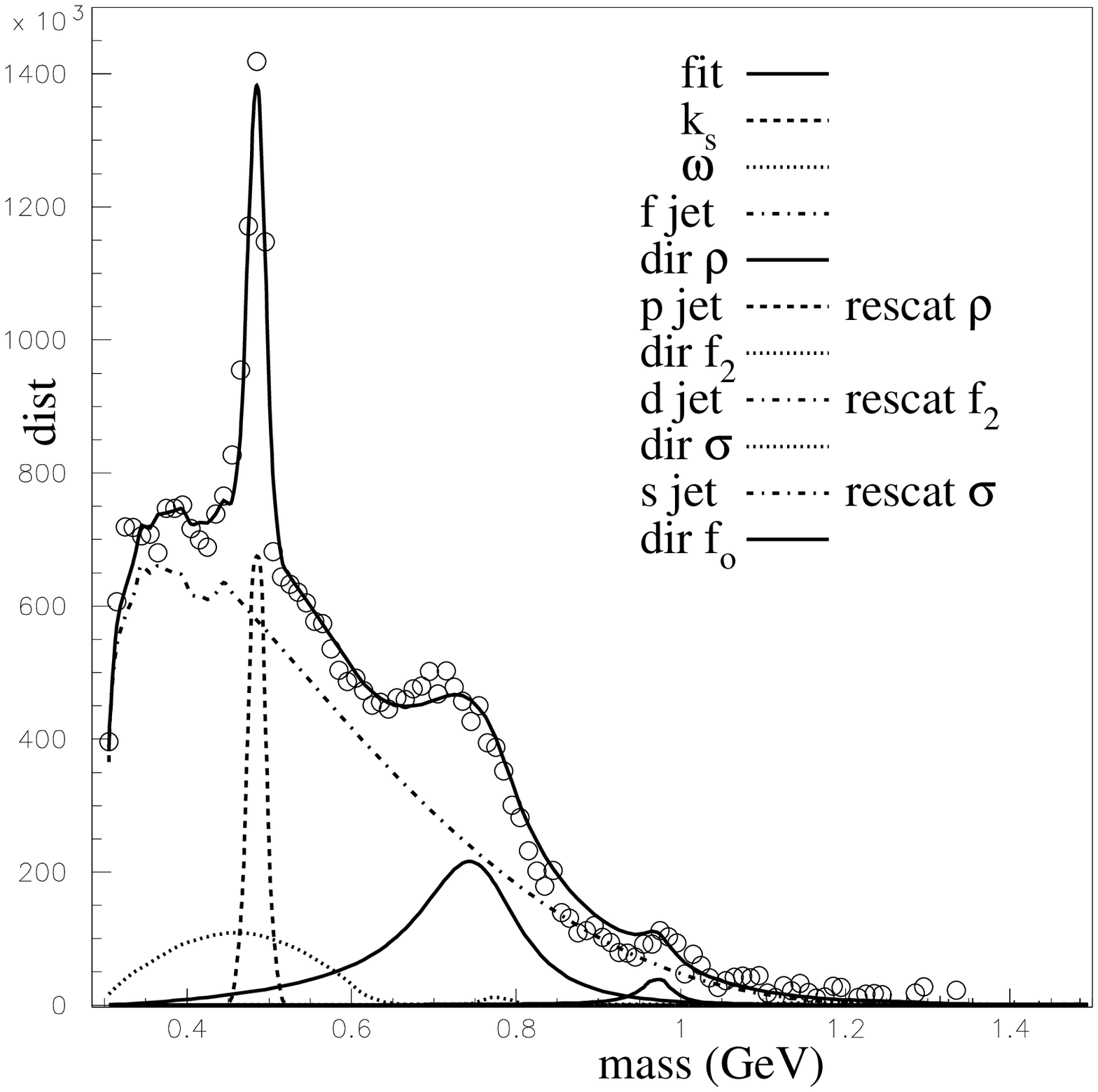}}
\end{center}
\vspace{2pt}
\caption{Fit to STAR dipion effective mass distribution (0.8 GeV/c $<$ $p_t$ 
$<$ 1.0 GeV/c) for Au-Au collisions at $\sqrt{s_{NN}} = $ 200 GeV 40\% to 80\% 
centrality using equation 7. See text for complete information.}
\label{fig4}
\end{figure}

\begin{figure}
\begin{center}
\mbox{
   \epsfysize 6.8in
   \epsfbox{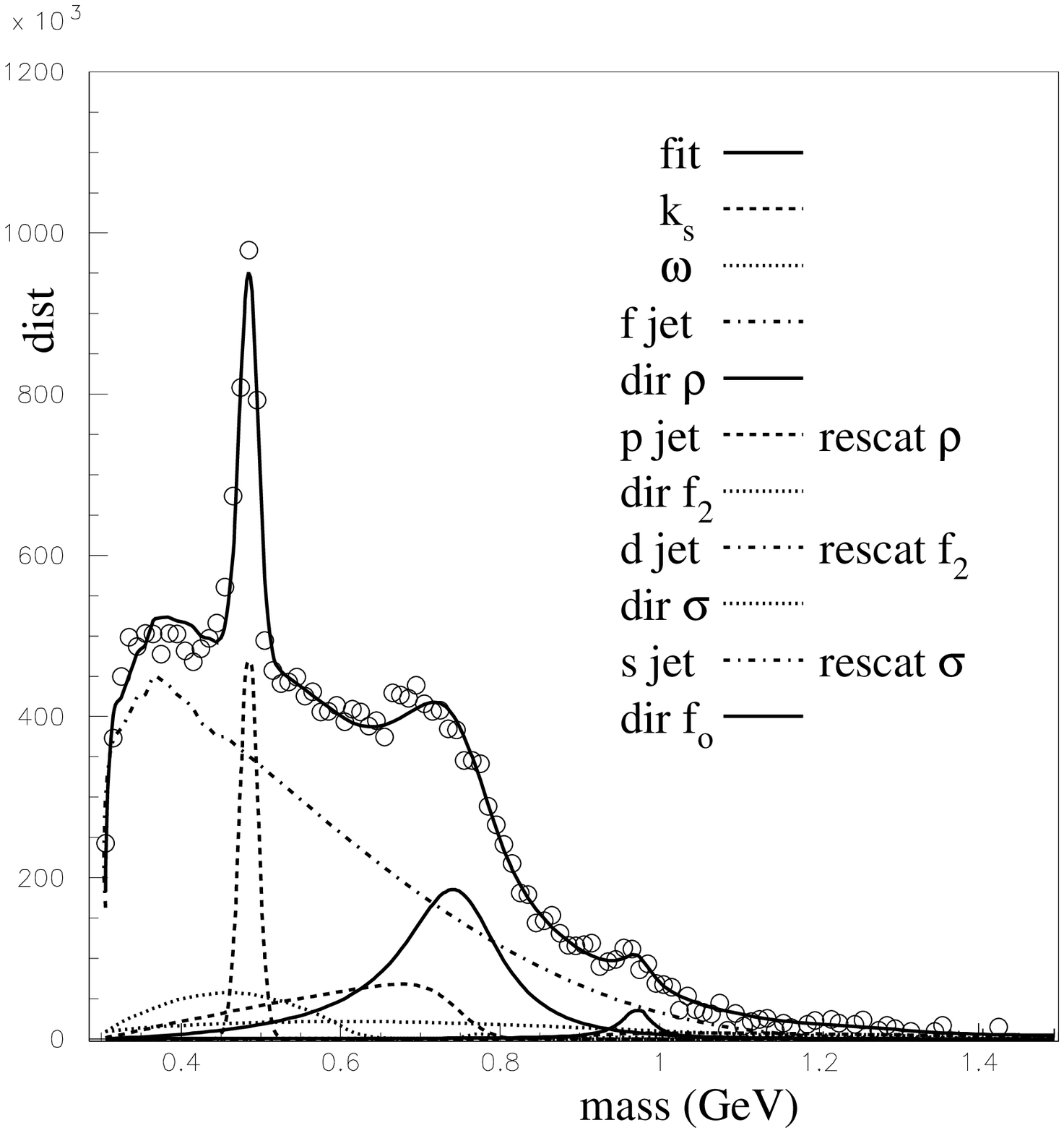}}
\end{center}
\vspace{2pt}
\caption{Fit to STAR dipion effective mass distribution (1.0 GeV/c $<$ $p_t$ 
$<$ 1.2 GeV/c) for Au-Au collisions at $\sqrt{s_{NN}} = $ 200 GeV 40\% to 80\% 
centrality using equation 7. See text for complete information.}
\label{fig5}
\end{figure}

\begin{figure}
\begin{center}
\mbox{
   \epsfysize 6.8in
   \epsfbox{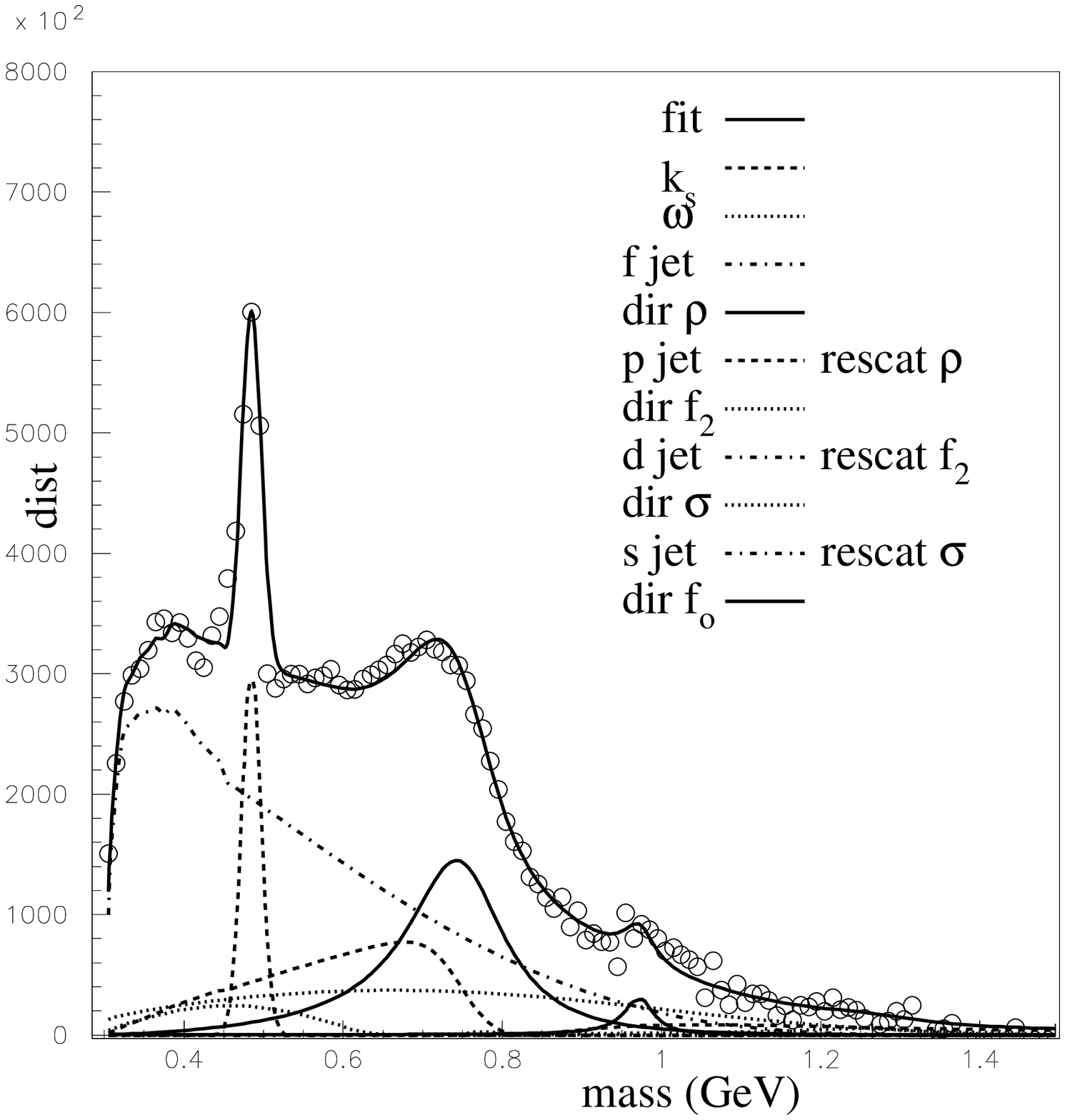}}
\end{center}
\vspace{2pt}
\caption{Fit to STAR dipion effective mass distribution (1.2 GeV/c $<$ $p_t$ 
$<$ 1.4 GeV/c) for Au-Au collisions at $\sqrt{s_{NN}} = $ 200 GeV 40\% to 80\% 
centrality using equation 7. See text for complete information.}
\label{fig6}
\end{figure}

\begin{figure}
\begin{center}
\mbox{
   \epsfysize 6.8in
   \epsfbox{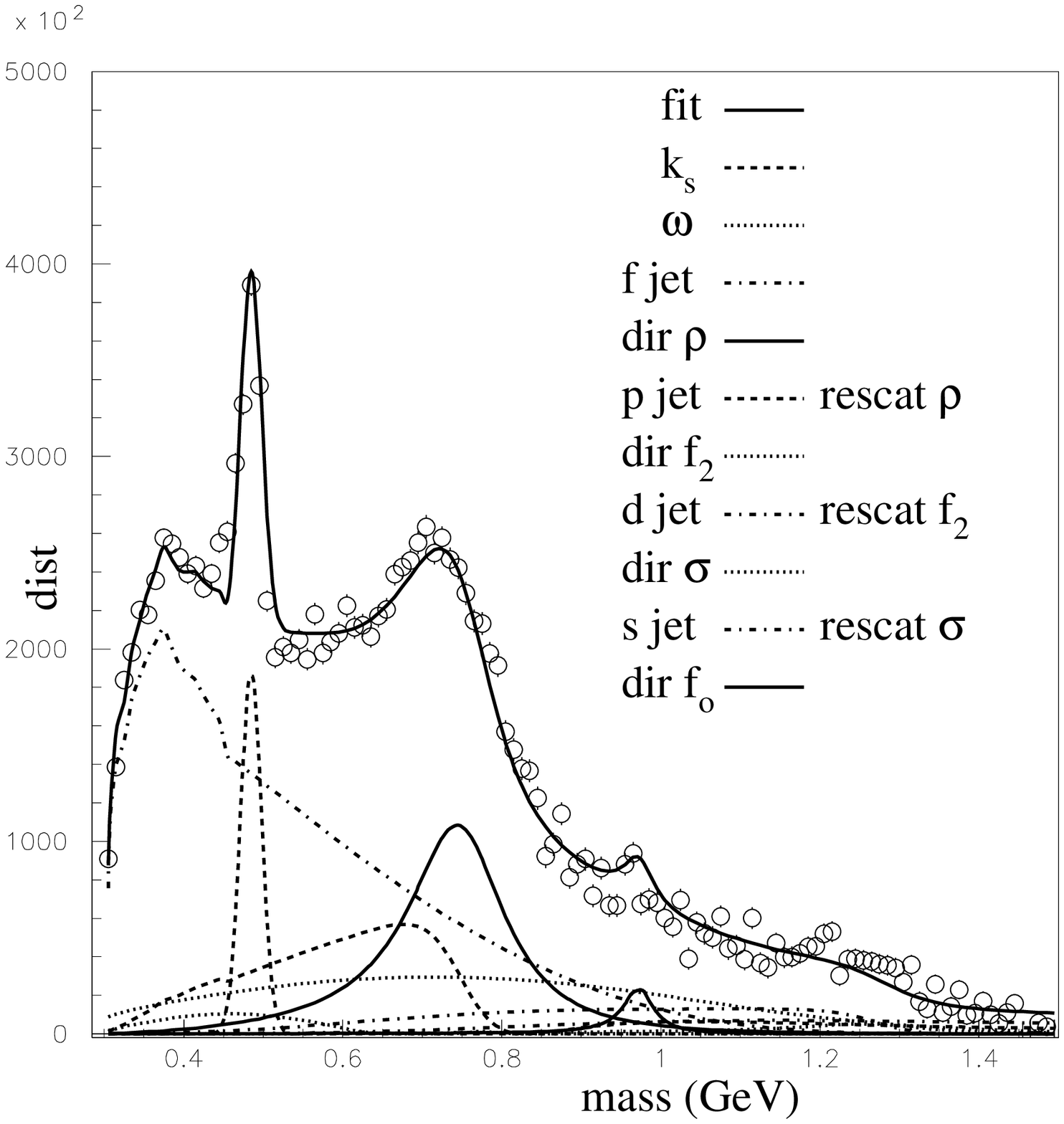}}
\end{center}
\vspace{2pt}
\caption{Fit to STAR dipion effective mass distribution (1.4 GeV/c $<$ $p_t$ 
$<$ 1.6 GeV/c) for Au-Au collisions at $\sqrt{s_{NN}} = $ 200 GeV 40\% to 80\% 
centrality using equation 7. See text for complete information.}
\label{fig7}
\end{figure}

\begin{figure}
\begin{center}
\mbox{
   \epsfysize 6.8in
   \epsfbox{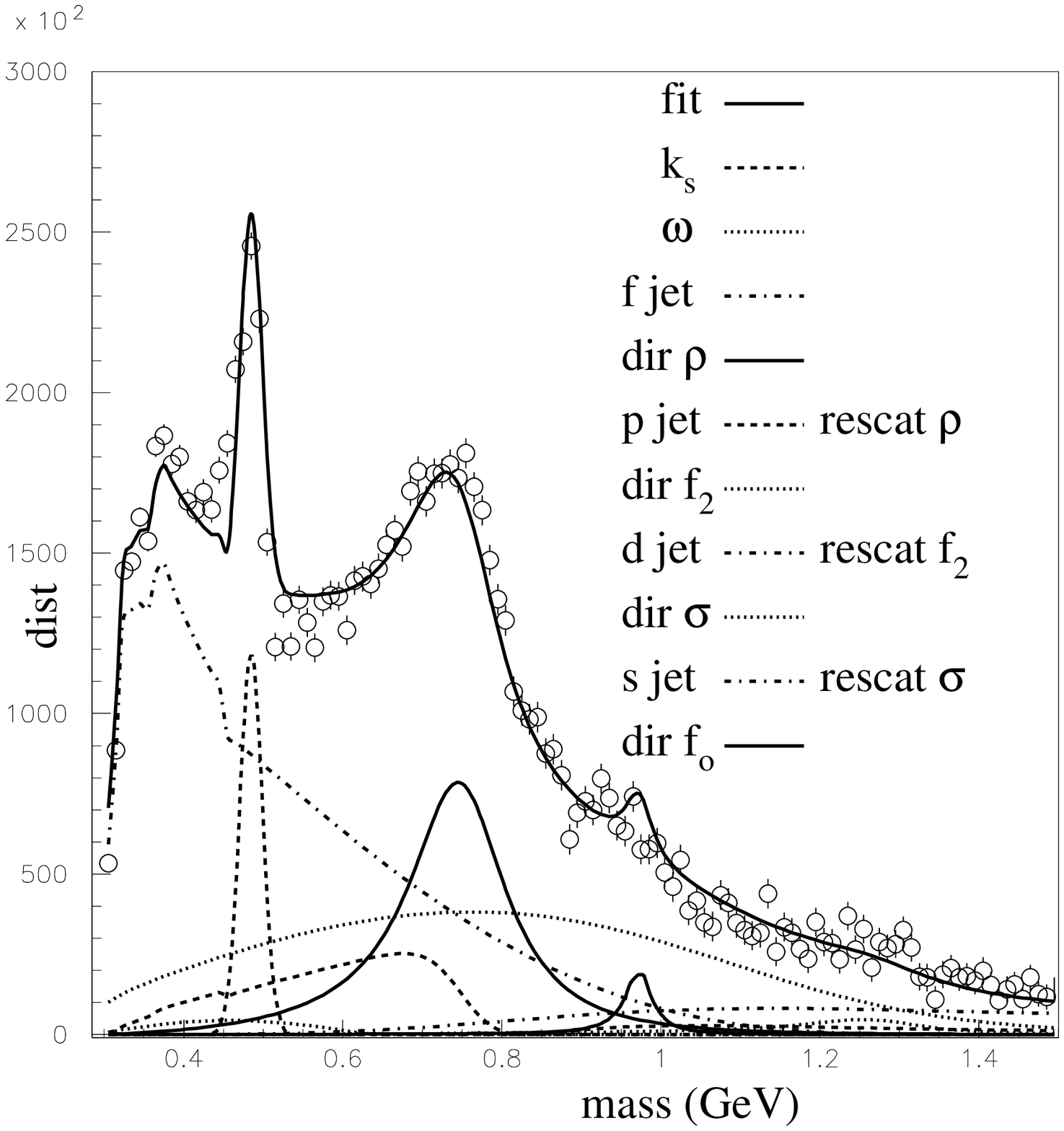}}
\end{center}
\vspace{2pt}
\caption{Fit to STAR dipion effective mass distribution (1.6 GeV/c $<$ $p_t$ 
$<$ 1.8 GeV/c) for Au-Au collisions at $\sqrt{s_{NN}} = $ 200 GeV 40\% to 80\% 
centrality using equation 7. See text for complete information.}
\label{fig8}
\end{figure}

\begin{figure}
\begin{center}
\mbox{
   \epsfysize 6.8in
   \epsfbox{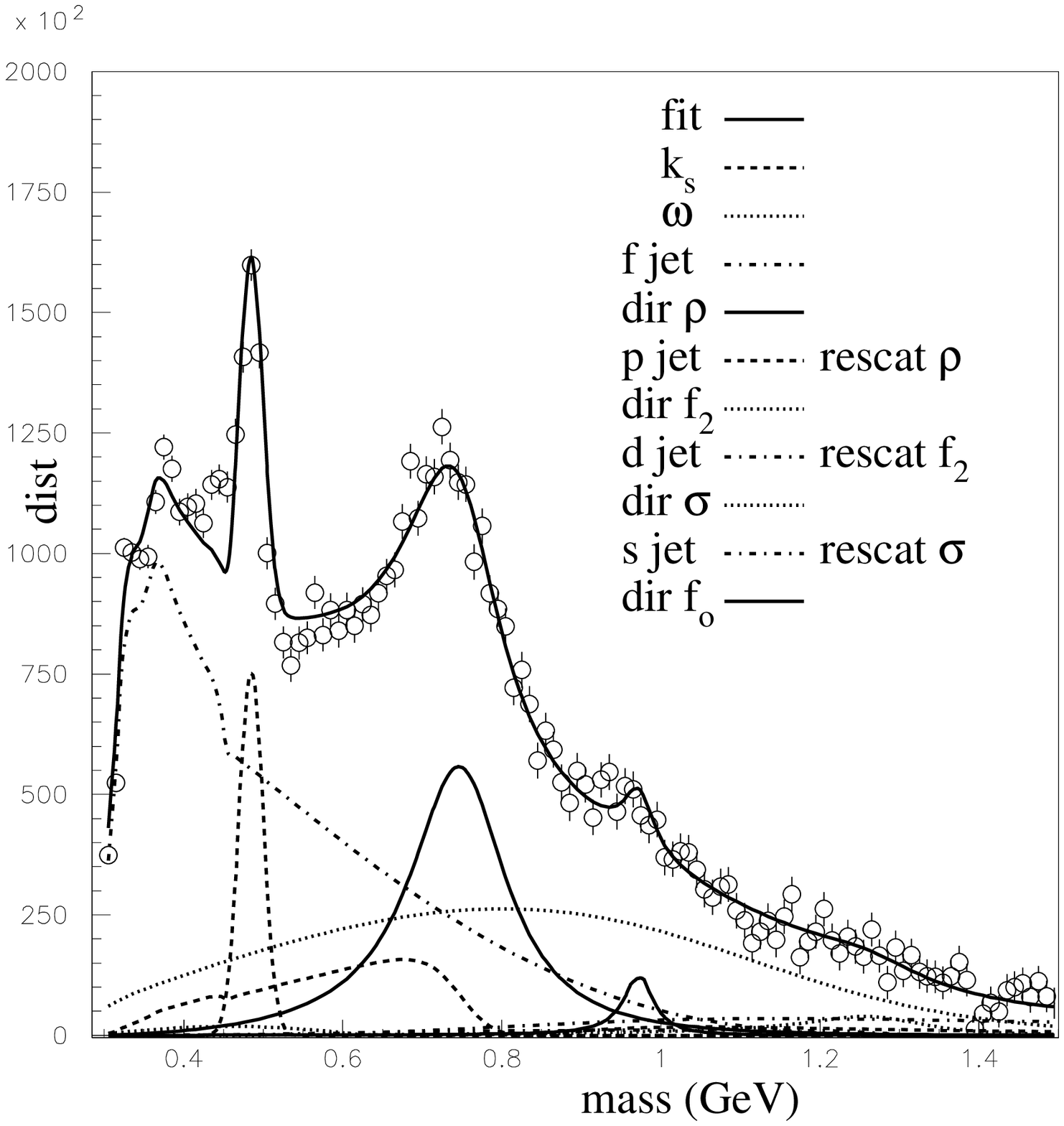}}
\end{center}
\vspace{2pt}
\caption{Fit to STAR dipion effective mass distribution (1.8 GeV/c $<$ $p_t$ 
$<$ 2.0 GeV/c) for Au-Au collisions at $\sqrt{s_{NN}} = $ 200 GeV 40\% to 80\% 
centrality using equation 7. See text for complete information.}
\label{fig9}
\end{figure}

\begin{figure}
\begin{center}
\mbox{
   \epsfysize 6.8in
   \epsfbox{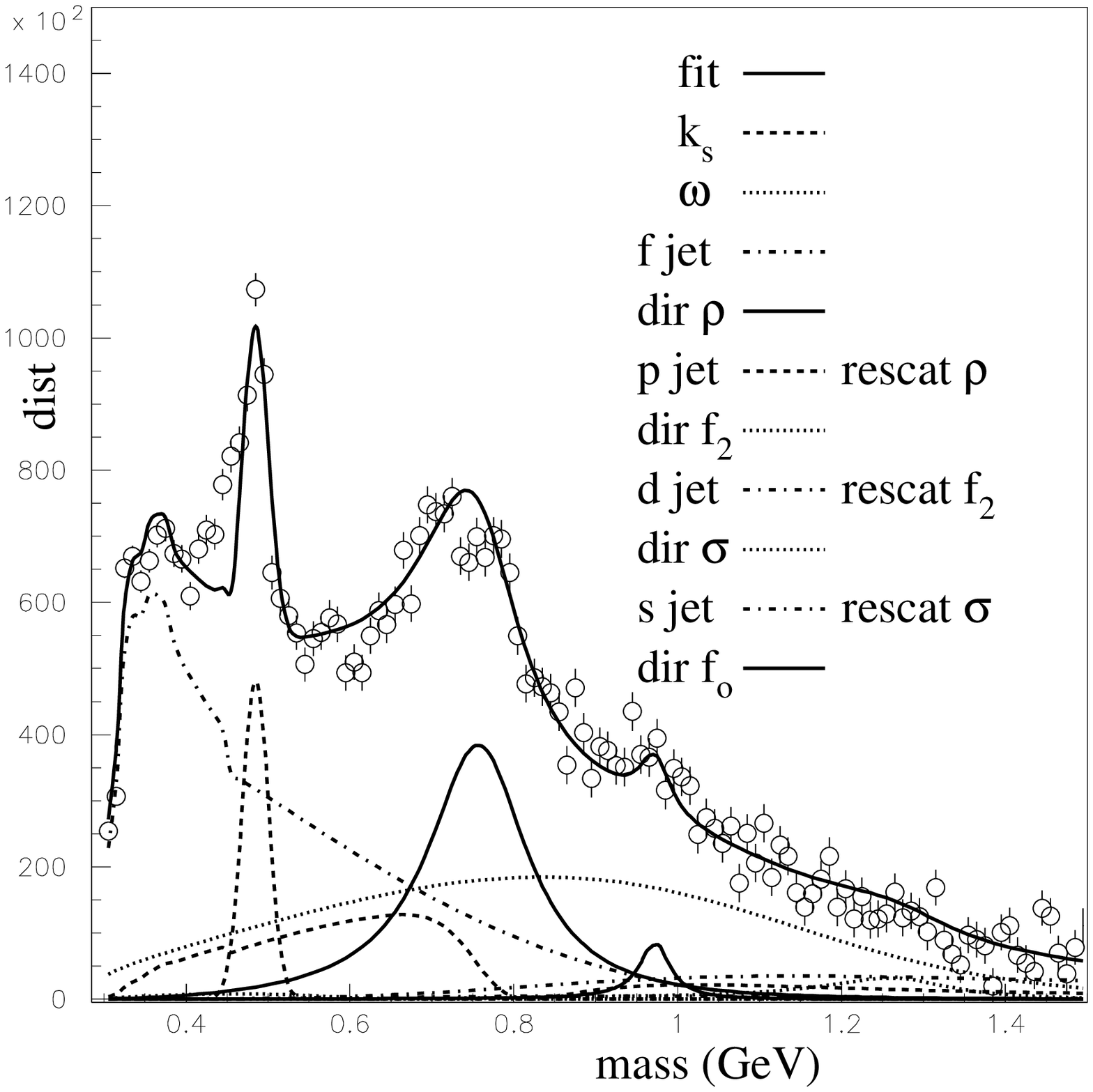}}
\end{center}
\vspace{2pt}
\caption{Fit to STAR dipion effective mass distribution (2.0 GeV/c $<$ $p_t$ 
$<$ 2.2 GeV/c) for Au-Au collisions at $\sqrt{s_{NN}} = $ 200 GeV 40\% to 80\% 
centrality using equation 7. See text for complete information.}
\label{fig10}
\end{figure}

\begin{figure}
\begin{center}
\mbox{
   \epsfysize 6.8in
   \epsfbox{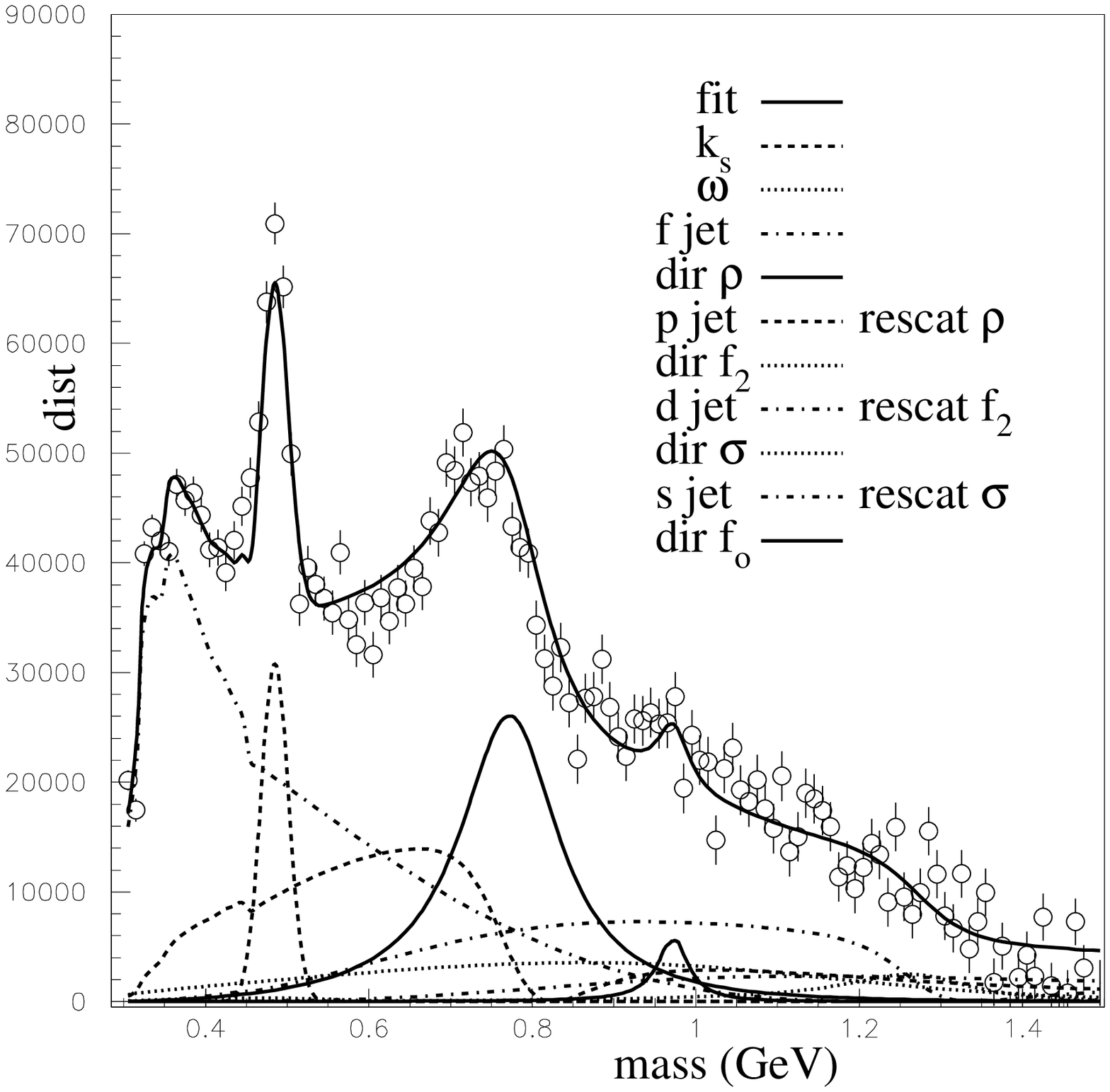}}
\end{center}
\vspace{2pt}
\caption{Fit to STAR dipion effective mass distribution (2.2 GeV/c $<$ $p_t$ 
$<$ 2.4 GeV/c) for Au-Au collisions at $\sqrt{s_{NN}} = $ 200 GeV 40\% to 80\% 
centrality using equation 7. See text for complete information.}
\label{fig11}
\end{figure}

\begin{figure}
\begin{center}
\mbox{
   \epsfysize 6.8in
   \epsfbox{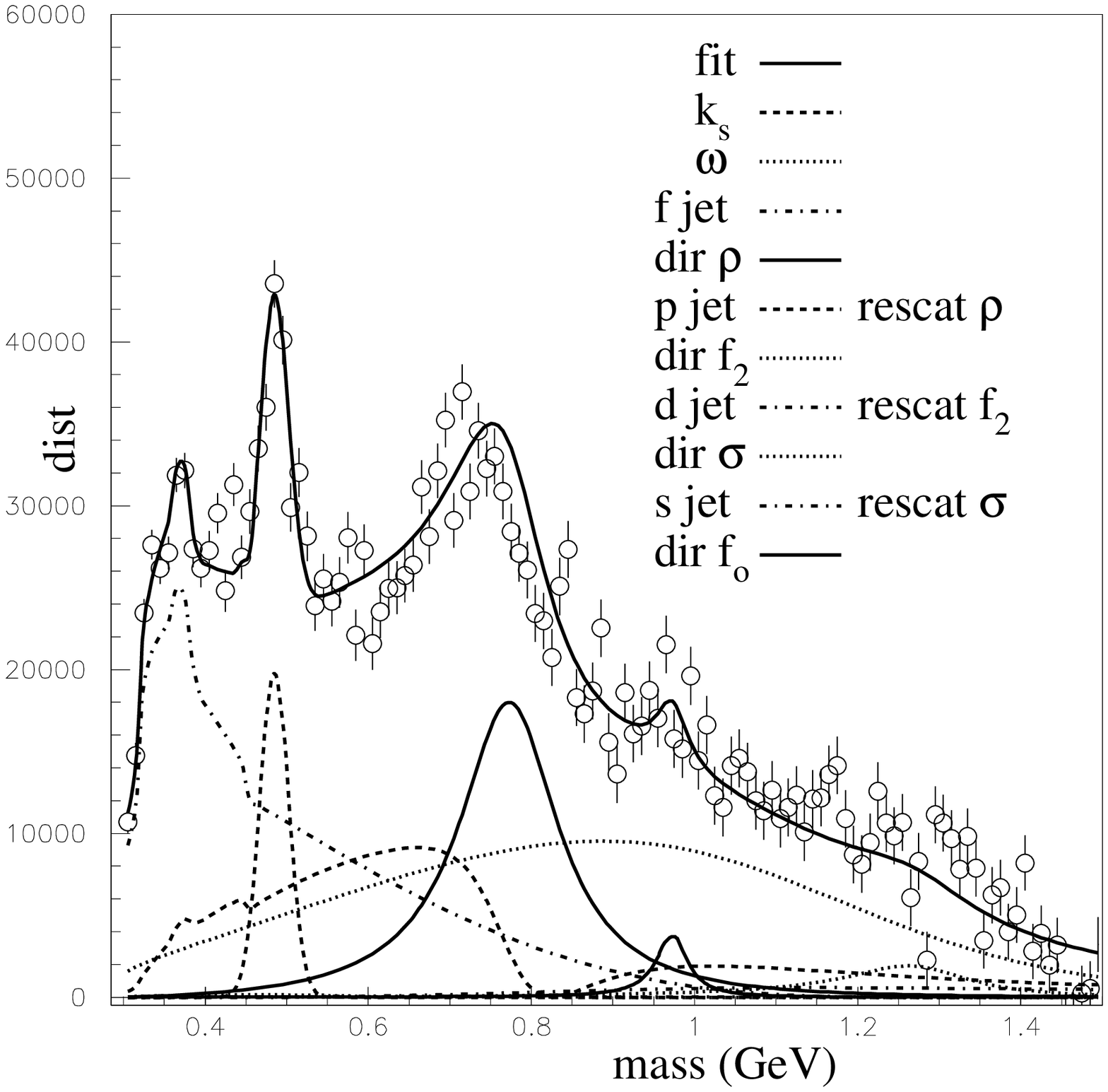}}
\end{center}
\vspace{2pt}
\caption{Fit to STAR dipion effective mass distribution (2.4 GeV/c $<$ $p_t$ 
$<$ 2.6 GeV/c) for Au-Au collisions at $\sqrt{s_{NN}} = $ 200 GeV 40\% to 80\% 
centrality using equation 7. See text for complete information.}
\label{fig12}
\end{figure}

\begin{figure}
\begin{center}
\mbox{
   \epsfysize 6.8in
   \epsfbox{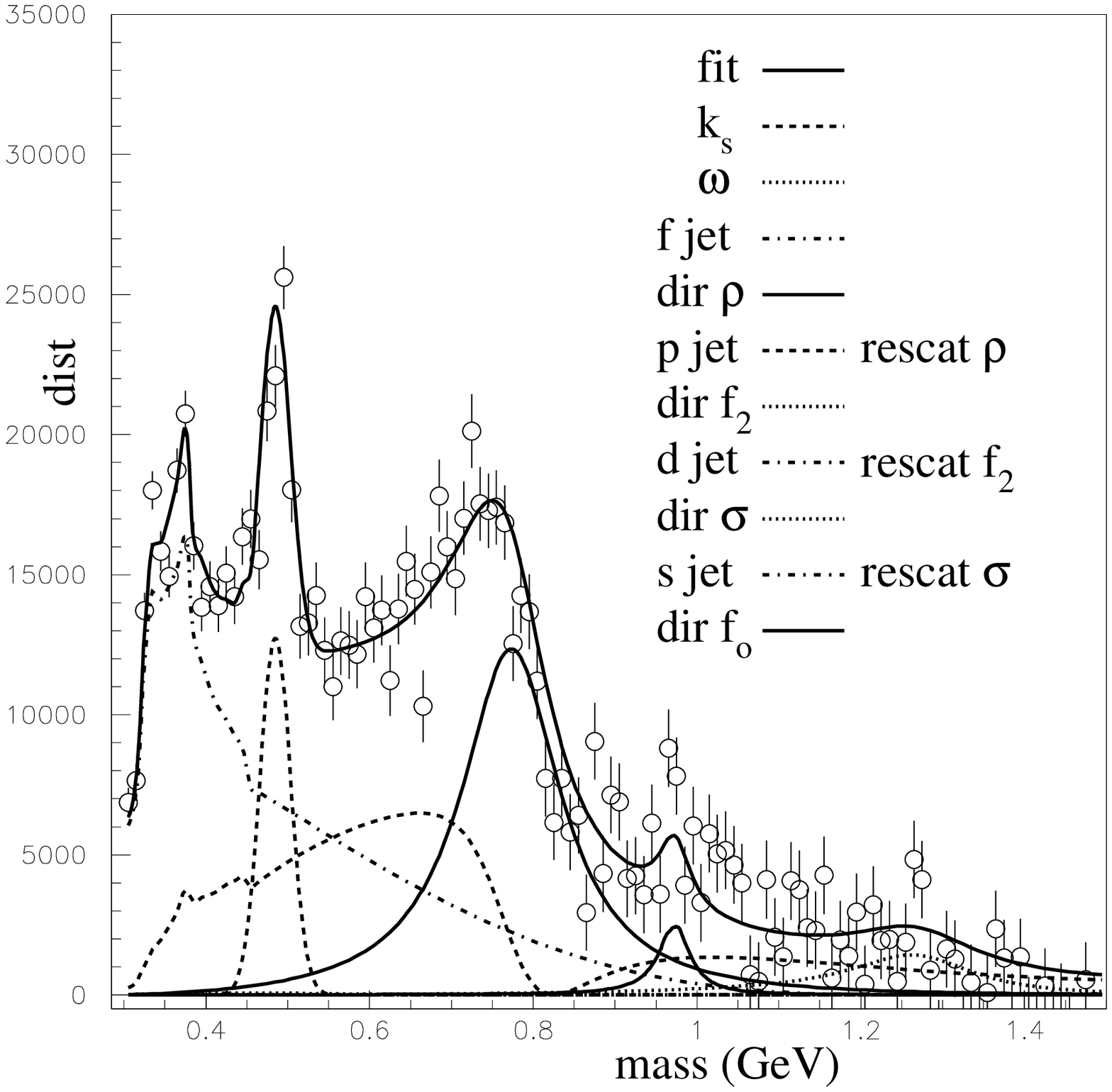}}
\end{center}
\vspace{2pt}
\caption{Fit to STAR dipion effective mass distribution (2.6 GeV/c $<$ $p_t$ 
$<$ 2.8 GeV/c) for Au-Au collisions at $\sqrt{s_{NN}} = $ 200 GeV 40\% to 80\% 
centrality using equation 7. See text for complete information.}
\label{fig13}
\end{figure}

\begin{figure}
\begin{center}
\mbox{
   \epsfysize 6.8in
   \epsfbox{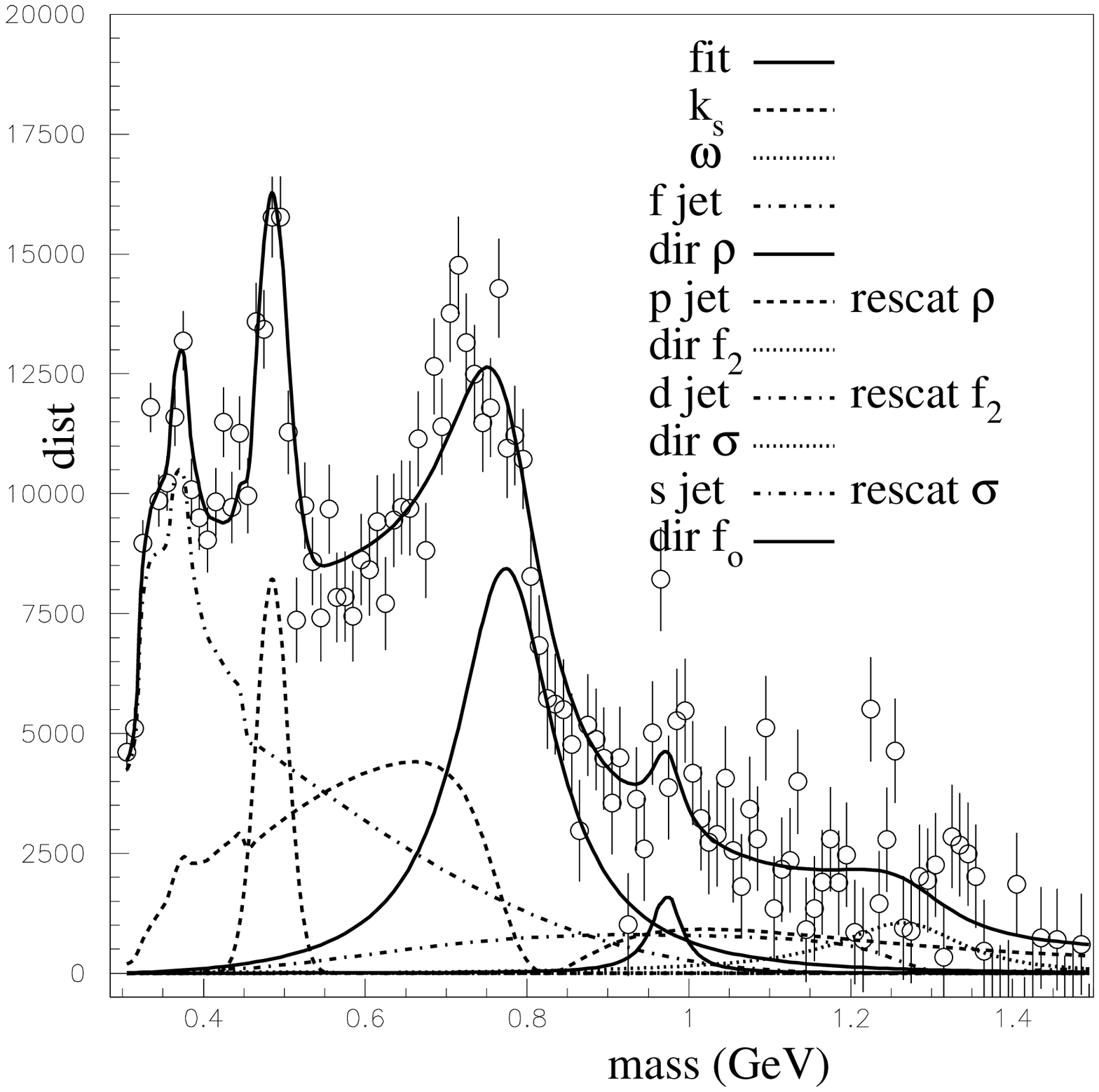}}
\end{center}
\vspace{2pt}
\caption{Fit to STAR dipion effective mass distribution (2.8 GeV/c $<$ $p_t$ 
$<$ 3.0 GeV/c) for Au-Au collisions at $\sqrt{s_{NN}} = $ 200 GeV 40\% to 80\% 
centrality using equation 7. See text for complete information.}
\label{fig14}
\end{figure}

\begin{figure}
\begin{center}
\mbox{
   \epsfysize 6.8in
   \epsfbox{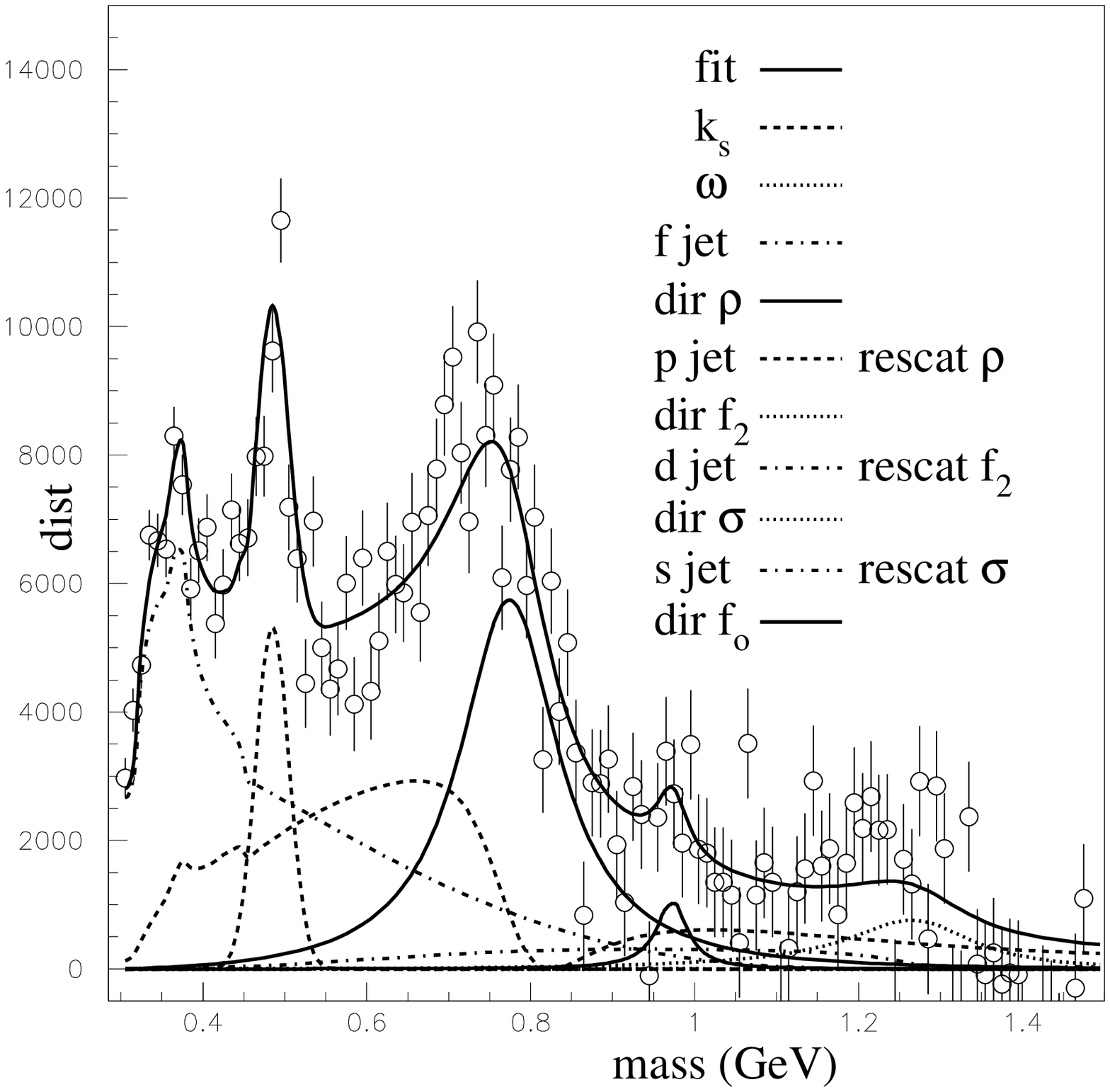}}
\end{center}
\vspace{2pt}
\caption{Fit to STAR dipion effective mass distribution (3.0 GeV/c $<$ $p_t$ 
$<$ 3.2 GeV/c) for Au-Au collisions at $\sqrt{s_{NN}} = $ 200 GeV 40\% to 80\% 
centrality using equation 7. See text for complete information.}
\label{fig15}
\end{figure}

\begin{figure}
\begin{center}
\mbox{
   \epsfysize 6.8in
   \epsfbox{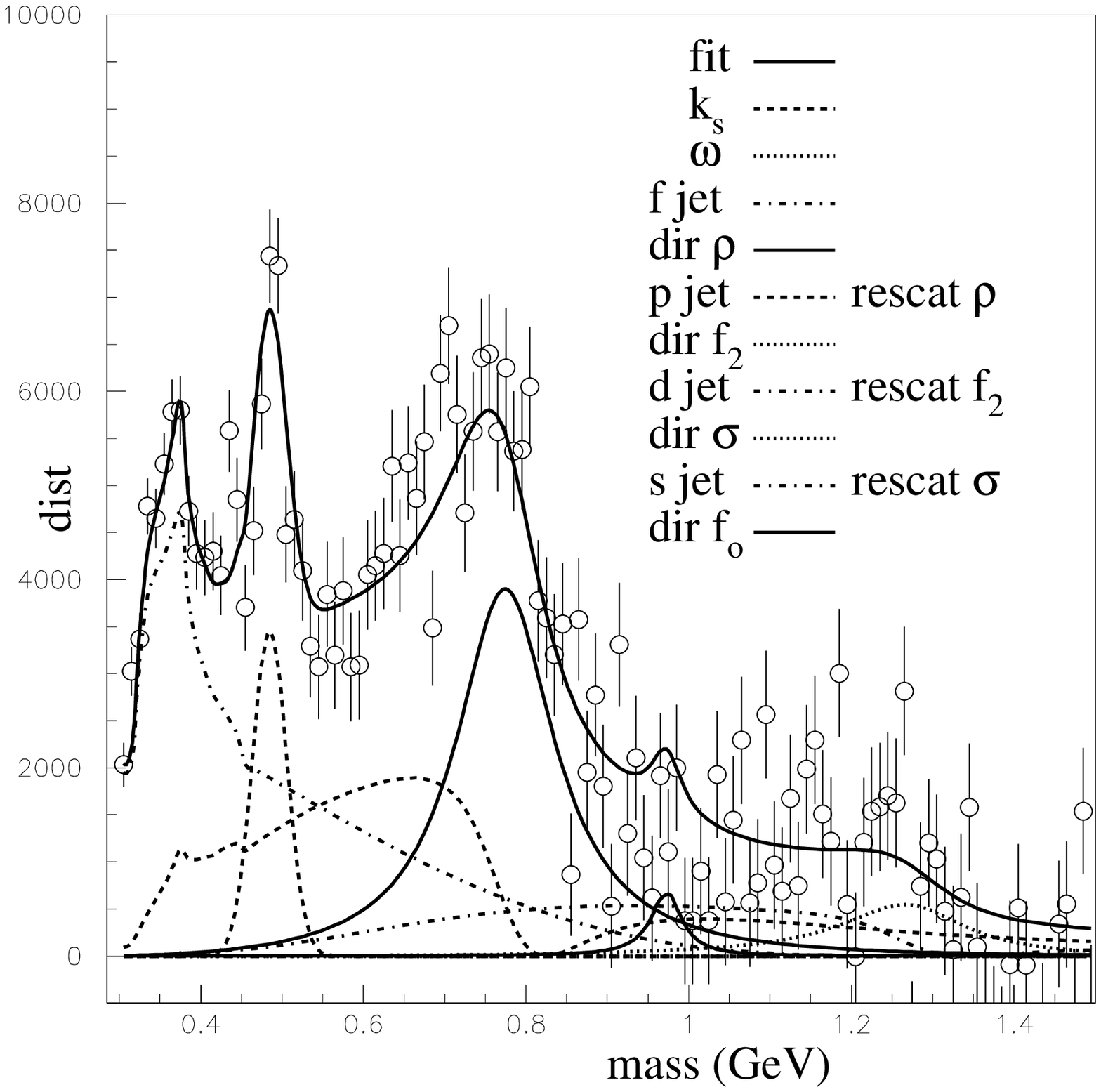}}
\end{center}
\vspace{2pt}
\caption{Fit to STAR dipion effective mass distribution (3.2 GeV/c $<$ $p_t$ 
$<$ 3.4 GeV/c) for Au-Au collisions at $\sqrt{s_{NN}} = $ 200 GeV 40\% to 80\% 
centrality using equation 7. See text for complete information.}
\label{fig16}
\end{figure}

\begin{figure}
\begin{center}
\mbox{
   \epsfysize 6.8in
   \epsfbox{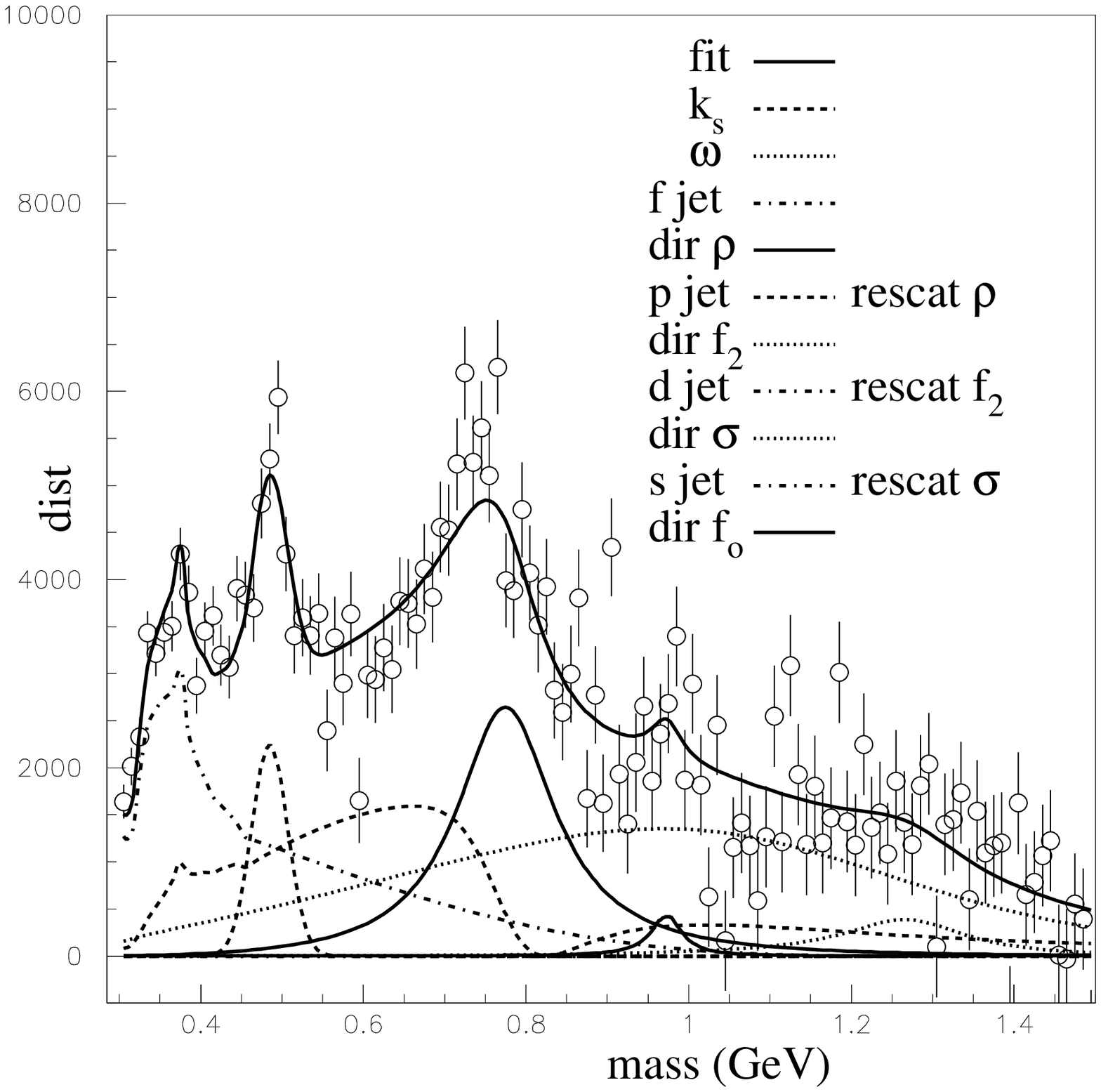}}
\end{center}
\vspace{2pt}
\caption{Fit to STAR dipion effective mass distribution (3.4 GeV/c $<$ $p_t$ 
$<$ 3.6 GeV/c) for Au-Au collisions at $\sqrt{s_{NN}} = $ 200 GeV 40\% to 80\% 
centrality using equation 7. See text for complete information.}
\label{fig17}
\end{figure}

\clearpage

\begin{figure}
\begin{center}
\mbox{
   \epsfysize 6.8in
   \epsfbox{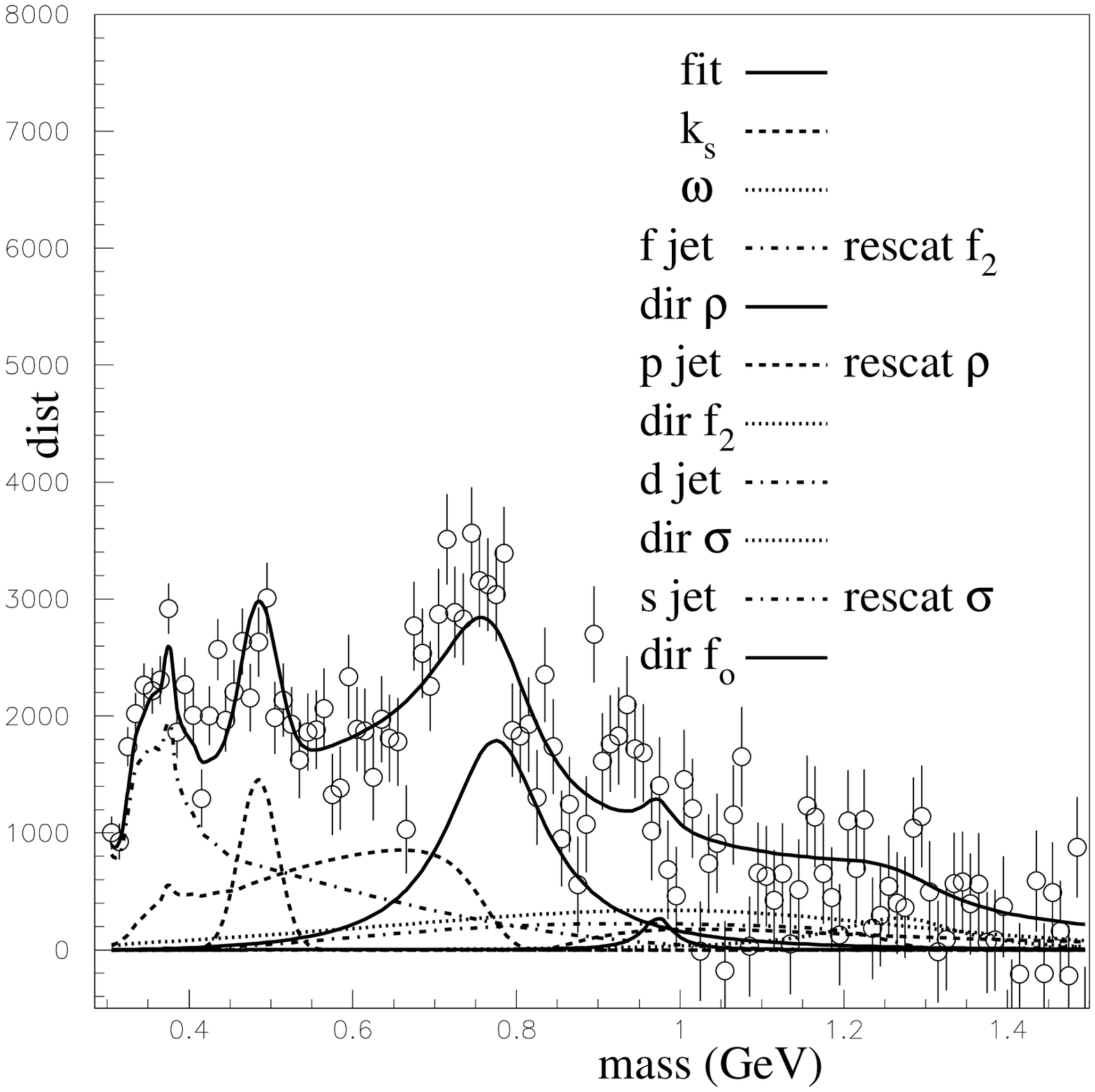}}
\end{center}
\vspace{2pt}
\caption{Fit to STAR dipion effective mass distribution (3.6 GeV/c $<$ $p_t$ 
$<$ 3.8 GeV/c) for Au-Au collisions at $\sqrt{s_{NN}} = $ 200 GeV 40\% to 80\% 
centrality using equation 7. See text for complete information.}
\label{fig18}
\end{figure}

\begin{figure}
\begin{center}
\mbox{
   \epsfysize 6.8in
   \epsfbox{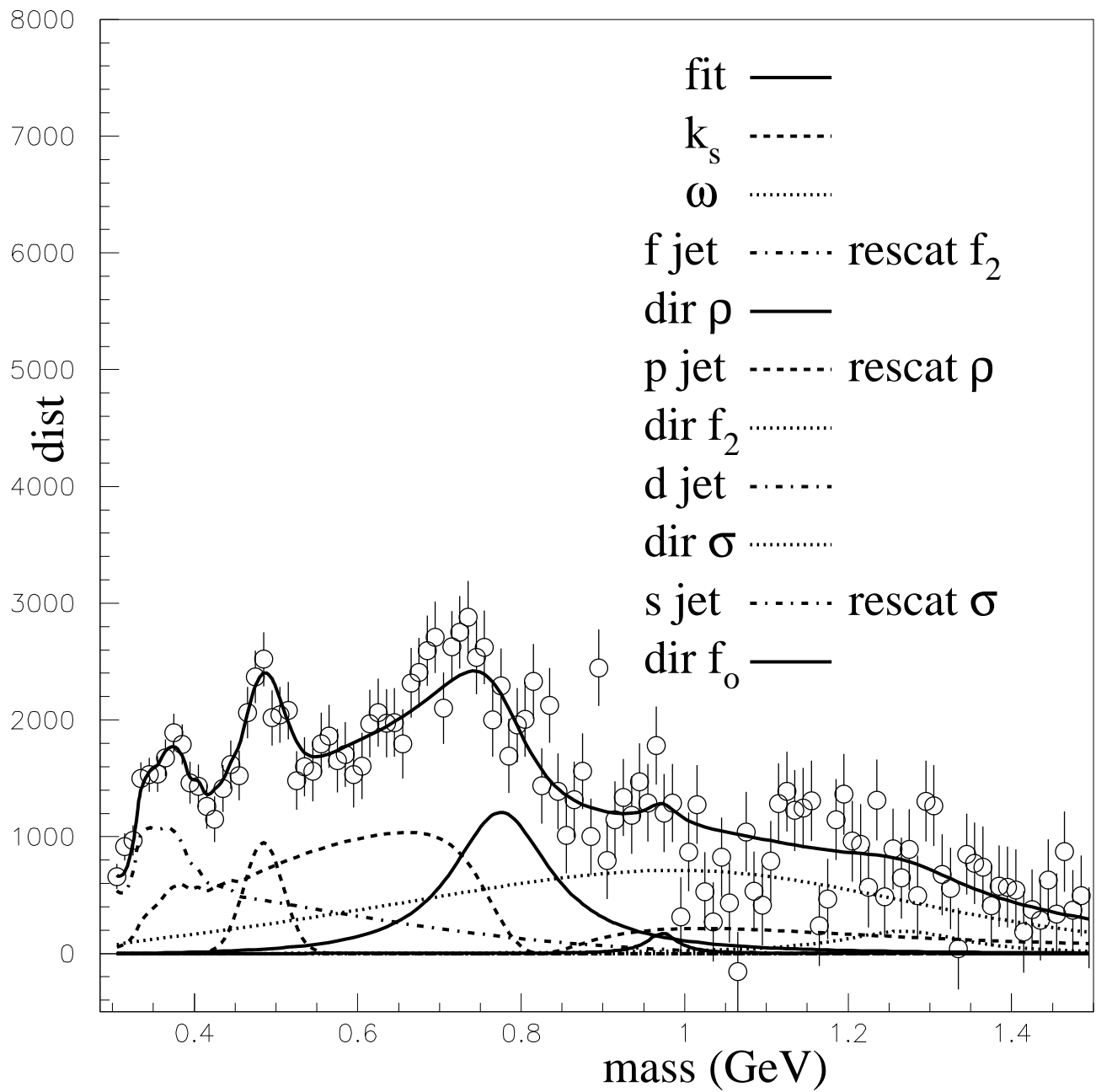}}
\end{center}
\vspace{2pt}
\caption{Fit to STAR dipion effective mass distribution (3.8 GeV/c $<$ $p_t$ 
$<$ 4.0 GeV/c) for Au-Au collisions at $\sqrt{s_{NN}} = $ 200 GeV 40\% to 80\% 
centrality using equation 7. See text for complete information.}
\label{fig19}
\end{figure}

\clearpage

The direct cross sectional yield of the thermally produced states into $\pi^+$
$\pi^-$ is shown in Figure 20. The states are $k_s$, $\rho$, $f_0$ and $f_2$
and the yields come from the exponential fits to the direct thermal component
of equation 7.

\begin{figure}
\begin{center}
\mbox{
   \epsfysize 6.8in
   \epsfbox{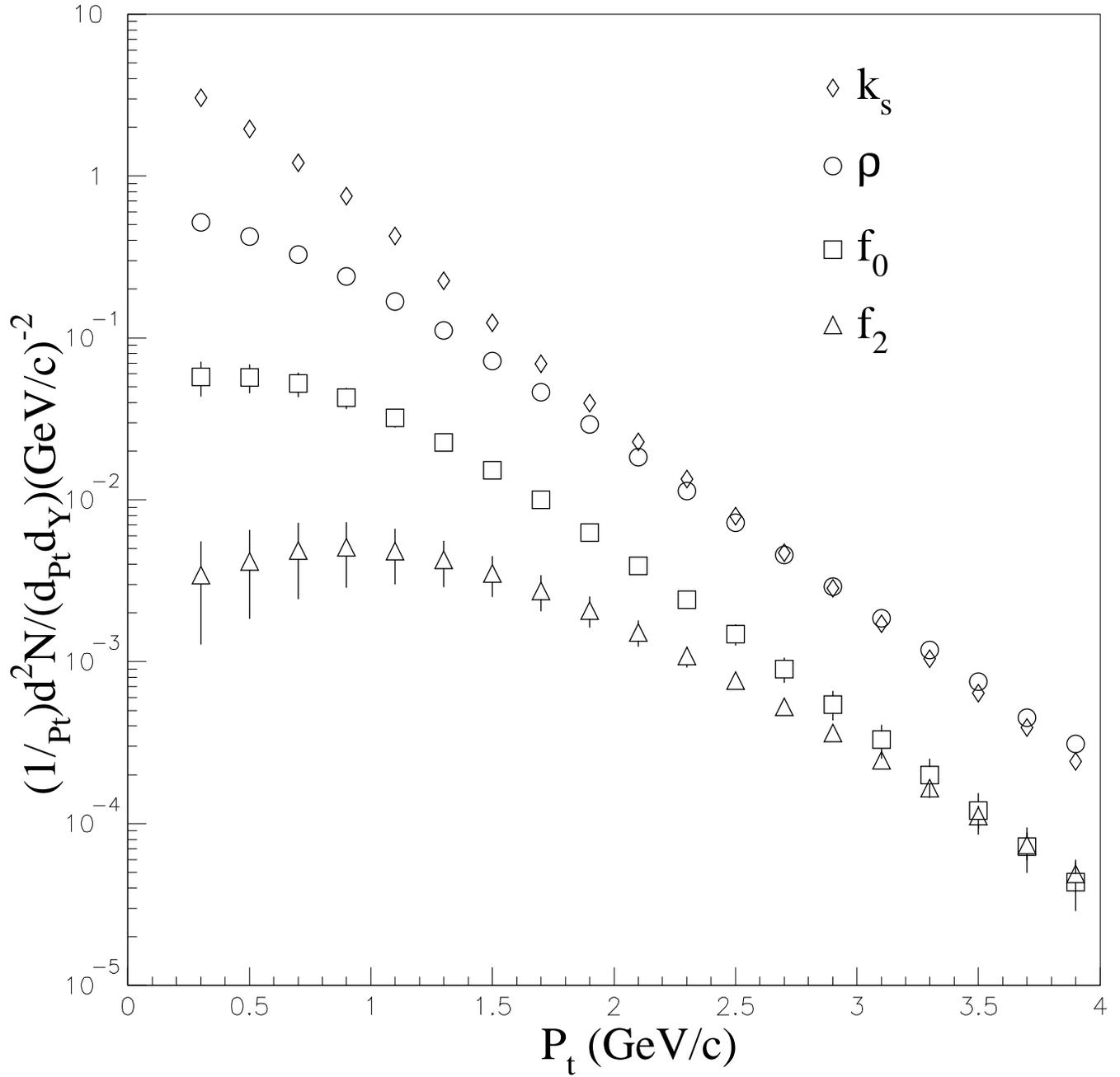}}
\end{center}
\vspace{2pt}
\caption{The direct thermal $p_t$ yeild into the $\pi^+ \pi^-$ channel as seen 
in the STAR dipion effective mass fit for the $k_s$, $\rho$, $f_0$ and $f_2$ 
using equation 7. See text for complete information.}
\label{fig20}
\end{figure}

\section{Summary and Discussion}
 The Dipion Cocktail Part 2 applies a formalism derived in The Dipion Cocktail 
Part 1 which considered a mixture of sources present in the dipion 
mass spectrum of the heavy ion fireball. Part 1 showed that both thermal or 
soft production of hadrons and the minijet fragmented hadrons can
be described through a set of unified formal equations. Part 2(this paper) 
applies this Part 1 formalism to the $p_t$ dependence of dipions for Au-Au 
collisions at $\sqrt{s_{NN}} = $ 200 GeV and 40\% to 80\% centrality. 

Part 1 started with the basic definition of elastic $\pi \pi$ 
scattering. Next showed how re-scattering of pions depends on the unitary 
condition that interactions present in the phase shift of an orbital state 
must interact all the time. The process of parton fragmentation into dipion 
states through unitarity leads to a equation of production and re-scattering 
in a given orbital quantum number. This equation(equation 7) has two 
components in each orbital state: one being the thermal production of 
resonances in a dipion orbital state, the other is the re-scattering of 
dipions coming from parton or minijet fragmentation into the dipion orbital 
state which do not come directly from the resonance. Unitarity requires that 
there most be re-scatter through resonance phase shifts which we defined 
through Breit-Wigner parameters (mass, width). 

We have fitted 19 dipion $p_t$ ranges(see Table I) using equation 7. We 
included minijets up to $\ell$ = 3 and resonances $\sigma$ $\ell$ = 0,
$\rho(770)$ $\ell$ = 1, and $f_2(1270)$ $\ell$ = 2. Using the arguments of 
Sec. 3 of Part 1, we added the $f_0$ as a direct thermal term ($|T_0|_1^2$) and
only the $\sigma$ interfered with $\ell$ = 0 minijet background. Two other 
thermal terms are present in the cocktail, the $k^0_s$ and the $\omega_0$. 
All the thermal terms have an exponential behavior with dipion $p_t$ and are
shown Figure 20. The spectrum of the minijet partial waves is obtained from
PYTHIA\cite{pythia}(see Sec. 5.2 of Part 1). We let the data determine which
minijet partial wave to add. We find only Swave minijet background is important
until $p_t$ equal to 1.1 GeV/c. Above 1.5 GeV/c all four minijet partial waves
are used up to Fwave. It should be noted Dwave and Fwave are small effects. 
We used PDG\cite{PDG} for the $f_2(1270)$ mass = 1.275 GeV and width = 
.185 GeV. The $f_0$ was fitted obtaining mass = 0.9727 $\pm$ .0039 GeV and
width = 0.04512 $\pm$ 0.01128 GeV. The $\sigma$ mass and width used was fixed
because it was ill determined. The mass used was mass = 1.011 GeV and width =
1.015 GeV. 

For the $\alpha$ parameter in $p_t$ bins up to 1.1 GeV/c the minijet Swave 
interference is the determining factor. Above 1.1 GeV/c the Pwave interference
becomes most important. The values of $\alpha$ which gives a reasonable fit are
shown in Table II.

We have determined that the $\sigma$ pole or Breit-Wigner parameters is so far
away from the real axis thus it is too short lived to be influenced by hadronic
interactions. The $\rho$ phase shift being of a life time comparable to 
hadronic interaction taking place becomes most sensitive. We have found as a
function of $\alpha$ the best of $\rho$ width is always 0.147 GeV with an error
of $\pm$ .007 GeV. The mass however decreases as $\alpha$ grows, reaching a 
minimum of 0.738 GeV at an $\alpha$ of 0.907. This is a mass shift of 37 MeV. 
An $\alpha$ of 0.504 is the smallest $\alpha$ we find in our fits. A mass of 
0.775 GeV is the best fit when the value of $\alpha$ is at 0.504. Using 
equation 9 in Table II we determine the radius of $\pi \pi$ re-scattering for 
each $p_t$ range. Table II shows an interesting density effect around dipion 
$p_t$ of 0.6 to 1.0 GeV/c. If one consider that $p_t$ maybe related to fireball
size through the idea of hubble flow, then pions with a $p_t$ of around 
0.4 GeV/c maybe coming from a less dense region in the central part of the 
fireball. This could be a density wave effect. 

\section{Acknowledgments}

This research was supported by the U.S. Department of Energy under Contract No.
DE-AC02-98CH10886. The author thanks William Love for the STAR analysis of the 
angular correlation data from Run 4. Also for his assistance in the production 
of figures. It is sad that he is gone.

\end{document}